\documentclass[preprint,5p]{elsarticle}

\usepackage{graphicx}
\usepackage{amsmath}
\usepackage{dcolumn}
\usepackage{color}
\usepackage{url}

\newcommand{\ket}[1]{\ensuremath{\left|#1\right\rangle}}

\journal{Physics Letters A}

\begin{document}

\begin{frontmatter}

\title{Quantum computation with classical light: \\implementation of the Deutsch-Jozsa Algorithm}

\author[itesm,wits,kwazulu1]{Benjamin Perez-Garcia}

\author[wits]{Melanie McLaren}

\author[kwazulu1,iqst]{Sandeep K. Goyal}

\author[itesm]{\\Raul I Hernandez-Aranda}

\author[wits]{Andrew Forbes}

\author[kwazulu1,kwazulu2]{Thomas Konrad\corref{cor1}}
\ead{konradt@ukzn.ac.za}

\cortext[cor1]{Corresponding author}

\address[itesm]{Photonics and Mathematical Optics Group, Tecnol\'ogico de Monterrey, Monterrey 64849, Mexico}

\address[wits]{University of the Witwatersrand, Private Bag 3, Johannesburg 2050, South Africa}

\address[kwazulu1]{School of Chemistry and Physics, University of KwaZulu-Natal, Private Bag X54001, Durban 4000, South Africa}

\address[kwazulu2]{National Institute of Theoretical Physics, Durban Node,  Private Bag X54001, Durban 4000, South Africa}

\address[iqst]{Institute of Quantum Science and Technology, University of Calgary, Alberta T2N 1N4, Canada}

\begin{abstract}
We propose an optical implementation of the Deutsch-Jozsa Algorithm using classical light in a binary decision-tree scheme. Our approach uses a ring cavity and linear optical devices in order to efficiently quarry the oracle functional values. In addition, we take advantage of the intrinsic Fourier transforming properties of a lens to read out whether the function given by the oracle is balanced or constant.
\end{abstract}


\end{frontmatter}

\section{Introduction}

Quantum computation has reached a stage wherein concepts and theory are properly understood while its implementation has not yet surpassed the  proof-of-principle level.  The difficulties lie in the state preparation and coherent control of a multitude of two-level systems carrying basic units of quantum information known as {\sl qubits}. In particular, decoherence due to the leak of information to the environment is a problem.  Even so, quantum computation promises unprecedented computational power and the possibility to tackle hitherto unsolvable computational tasks. This is done by simultaneously processing a multitude of numbers encoded in large superpositions of the corresponding states of quantum systems. There are several quantum algorithms that have been experimentally realised. For instance, the Deutsch \cite{Deutsch1985} and Deutsch-Jozsa \cite{Deutsch1992} algorithms have been implemented in nuclear-magnetic-resonance systems \cite{Jones1998,Linden1998,Dorai2000,Kim2000,Mahesh2001,Mangold2004}, QED cavities \cite{Zheng2004,Rong-Can2006,Hong-Fu2008,Wang2009}, quantum dots \cite{Bianucci2004,Scholz2006}, trapped ions \cite{Gulde2003}, light shifts\cite{Dasgupta2005}, superconducting quantum processors \cite{Zheng2008}, nitrogen-vacancy defect center \cite{Shi2010} and quantum optical systems \cite{Oliveira2005,Marques2012,Zhang2010,Zhang2012}. However, due to the aforementioned problems, the maximum number of qubits used in such implementations has not yet exceeded a few. For example, the greatest number of qubits used  for the Deutsch-Jozsa Algorithm was four \cite{Schuch2002}. Thus there has not yet been a computational problem solved on a quantum computer which was inaccessible for classical computers (Turing machines).

Here we propose to the best of our knowledge the first scalable implementation of the Deutsch-Jozsa Algorithm. A salient feature of our proposal is the use of the spatial degree of freedom of classical light fields, allowing to efficiently encode an unlimited number of qubits. After the Deutsch Algorithm \cite{PerezGarcia2015}, this is the second scheme we present of quantum computation with classical light and the first one that can outperform any Touring Machine.

\section{Deutsch-Jozsa algorithm}

The Deutsch-Jozsa Algorithm aims to distinguish between binary functions $f: \{0, 1, \ldots, N-1\} \rightarrow \{0,1\}$ that are constant, $f(x)=0$ or $f(x)=1$ for all arguments $x$,  and balanced functions $f$ which assume for $N/2$ of their arguments the value $0$ and for the other $N/2$ of arguments the value 1. The problem becomes intractable on Turing machines if the number $N$ is very large, the reason lies in the need to inquire up to $N/2 +1$ functional values from an oracle (or databank). For example, let the arguments of the function be stored in an input register of 200 Bits and run over the full possible range from $0$ to $2^{200} \approx 10^{60}$. If every query for a functional value would just take a pico second then evaluating $N/2 +1$ values would require up to $10^{47}$ s, a large multiple of the age of the universe. A quantum computer can access and process all the functional values at the same time based on the superposition principle and deterministically yield the right result using interference between of the processed states representing the functional values. 

The present implementation scheme uses a  version of the  Deutsch-Jozsa Algorithm \cite{Collins1998} which works only with  a single register with $n=\log_2 N$ qubits as explained in the following.  Initially each of $n$ qubits in  the input register prepared in state $\ket{0}$,  is processed by a  Hadamard transformation,  $\ket{0} \xrightarrow{H} \frac{1}{\sqrt{2}}(\ket{0}+\ket{1})$, yielding a superposition of states, which corresponds to  $2^n$ arguments $x$:
\begin{align}\label{eq:superposition}
&\ket{\Psi_{\text{in}}} = \ket{0,0,\ldots,0} \nonumber \\
&\xrightarrow{H^{\otimes n}} \frac{1}{\sqrt{2^n}}\sum_{x_1,\ldots ,x_n=0}^{1} \ket{x_1, \ldots x_n}= \frac{1}{\sqrt{2^n}}\sum_{x=0}^{2^n-1} \ket{x}.
\end{align}
Here $ x_1, \ldots x_n$ are the binary digits representing the number $x$.
Thereafter the value $0$ (or $1$) of $f(x)$ is encoded as relative phase factors $+1$ ($ \mbox{or}\, -1$) of the input state $\ket{x}$:   
\begin{equation}
\frac{1}{\sqrt{2^n}}\sum_{x=0}^{2^n-1} \ket{x} 
\xrightarrow{U_f} \frac{1}{\sqrt{2^n}}\sum_{x=0}^{2^n-1} (-1)^{f(x)} \ket{x}.\label{eq:transformation}
\end{equation}
Because the transformation  $\ket{x} \xrightarrow{U_f} (-1)^{f(x)}\ket{x}$ is linear it imprints the relative phases carrying the information about $f(x)$ for all  $x$-values in the superposition $\sum \ket{x}$ at the same time.  
A second Hadamard transformation of the $n$ qubits 
\begin{equation}\label{eq:outputquantum}
 \frac{1}{\sqrt{2^n}}\sum_{x=0}^{2^n-1} (-1)^{f(x)} \ket{x}
\xrightarrow{H^{\otimes n}} \frac{1}{{2^n}} \sum_{x=0}^{2^n-1}\sum_{z=0}^{2^n-1}(-1)^{x\cdot z + f(x)}\ket{z},
\end{equation}
yields the result of the computation as an interference effect. The term $x\cdot z$ denotes the bitwise inner product of $x$ and $z$.  If the function is constant, the second Hadamard transformation reverses the first one and propagates the system back into its initial state $\ket{0,0,\ldots,0}$. This can be interpreted as constructive interference of the equal phase factors forming the probability amplitude of state $\ket{0,0,\ldots,0}$, cp.\ right-hand side of  (\ref{eq:outputquantum}).  On the other hand, a balanced function results in the same number of positive ($+1$) and negative phase factors ($-1$) which destructively  interfere to a zero probability amplitude for the state  $\ket{0,0,\ldots,0}$. Whether the function is constant or balanced can thus be determined by measuring if  $\ket{0,0,\ldots,0}$ is populated or a state in its orthogonal complement. 

\begin{figure}[htp]
\centering
\includegraphics[width=80mm]{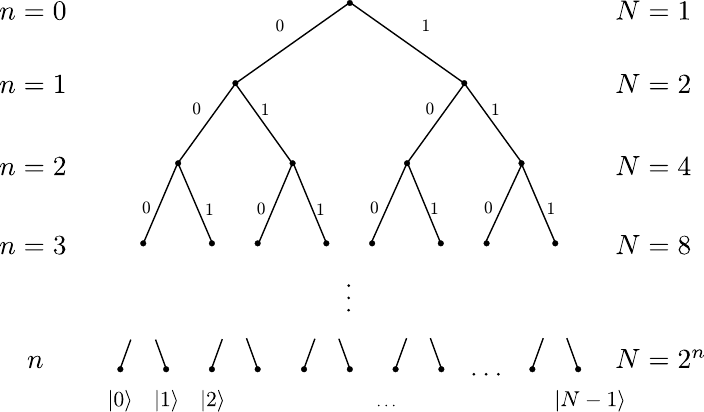}
\caption{Diagrammatic representation of scheme to prepare a superposition of $2^n$ states in $n$ steps. It resembles the quantum walk on a graph with two edges at each vertex (a binary tree). }
\label{fig:DiagEf}
\end{figure}

The central transformation (\ref{eq:transformation}), which encodes all functional values as binary phases for each argument $x$ into the register, has an immediate optical analog -- the imprint of position-dependent phases on the electric field on a transversal plane using a phase mask. A corresponding optical implementation of the Deutsch-Jozsa algorithm \cite{Puentes2004} realised the phase imprint by means of a spatial light modulator (SLM) which acts at a transversal plane on plane-wave laser light. The subsequent Hadamard transformation (\ref{eq:outputquantum}) is realised by means of the Fourier-transform properties of a thin lens (see \ref{app:fourier}). However, such a straight-forward encoding is based on the unary representation of each number $x$ corresponding to a position on the transversal plane, i.e., the phase shifts are encoded on the SLM argument by argument. This procedure requires therefore as many steps to encode functional values in form of phase shifts as there are arguments $x$. Hence the function cannot be written efficiently into the register and the solution fails, if the number of arguments is large. 

\section{Binary tree design}

\begin{figure}[htp]
\centering
\includegraphics[width=85mm]{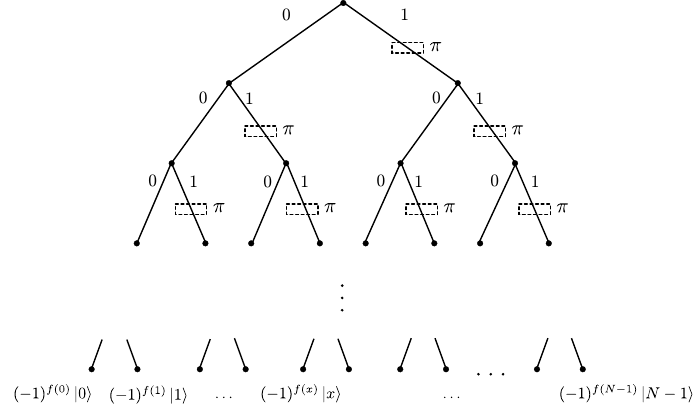}
\caption{Any binary function $f$ with $N=2^n$ arguments $x$  can be encoded  in the superposition $\sum (-1)^{f(x)}\ket{x}$ by means of individual $\pi$ phase shifts in the branches labeled "1". Moreover, all phase shifts on a given level of the tree can be implemented simultaneously by a single operation in the realisation scheme.}
\label{fig:DiagEfPhase}
\end{figure}

\begin{figure}[htp]
\centering
\includegraphics[width=85mm]{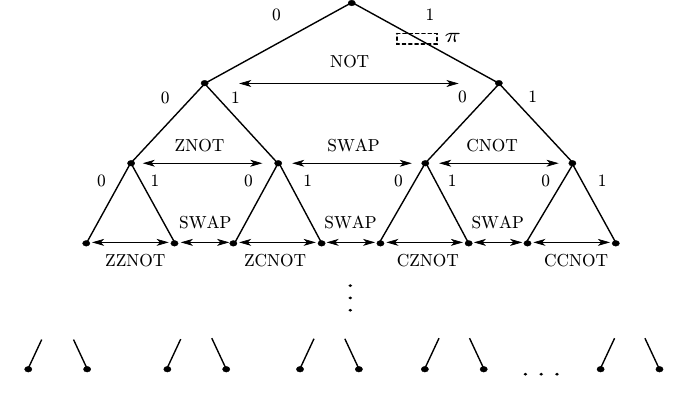}
\caption{Swapping states $\ket{0}$ and $\ket{1}$ on different levels of the binary tree corresponds to Not and controlled Not operations at different depth. For example, a swap on the third level given by $\ket{110}\leftrightarrow \ket{111}$  corresponds to a CCNOT operation, also know as Toffoli gate. Each balanced function can be  encoded by a specific combination of these swaps and a $\pi$ phase shift.}
\label{fig:DiagEfCnot}
\end{figure}

There are several ways to model quantum computation, for instance, the quantum circuit model and  measurement-based quantum computation \cite{Nielsen2000,Raussendorf2001}. Although binary tree designs have been previously reported \cite{Farhi1998, Kotiyal2014}, here we propose a novel binary decision tree design as a new paradigm to model quantum algorithms which can be helpful to provide an efficient classical optics picture of such algorithms (see Figs.\ \ref{fig:DiagEf}--\ref{fig:DiagEfCnot}). The method we present here combines the efficient creation of the superposition of all arguments $x\in\{0, 1, \ldots, 2^n-1\}$ with the encoding of the functional values in the form of phase factors. It is based on the optical approach of Daniela Dragoman \cite{Dragoman2002} to prepare a superposition of $2^n$ states $\ket{x}$  efficiently, i.e., in $n$ steps using a binary decomposition of the number $x$. Our approach modifies Dragoman's idea in order to implement the Deutsch-Jozsa algorithm by means of path qubits and a ring resonator.

The principle of the method is sketched in Fig.\ \ref{fig:DiagEf}. In each step of the preparation corresponding to a round trip in the ring cavity, one additional path qubit is generated by splitting each of the present light beams, i.e., doubling the number of beams. In the first step, a single path qubit is created  
\begin{equation}
\ket{\psi_{\text{in}}}=\ket{0}\rightarrow \frac{1}{\sqrt{2}}(\ket{0}+\ket{1})\,,
\label{inputtrafo0}
\end{equation}
where $\ket{0} (\ket{1})$ represents the left (right) part of the beam (cp.\ Fig.\ \ref{fig:DiagEf}).  
In the second step, each of the beams is split again:   
\begin{equation}
\frac{1}{\sqrt{2}}(\ket{0}+\ket{1})\rightarrow\frac{1}{2}(\ket{00}+\ket{01} +\ket{10}+\ket{11})\,.
\end{equation}
We denote by $\ket{x_1, x_2 }\equiv\ket{x_1}\ket{x_2}$ the part of the light beam that propagates in direction $x_1\in \{0,1\}$  and  $x_2\in \{0,1\}$  after the first and  second step, respectively.
In $n$ steps the state thus changes according to 
\begin{equation}
\ket{\psi_{\text{in}}}\rightarrow \frac{1}{\sqrt{2^n}} \sum_{x_1, \ldots, x_n= 0,1} \ket{x_1, \ldots x_n}.
\label{inputtrafo}
\end{equation} 
This procedure generates $2^n$ localised light fields, represented by the state $\ket{0}+\ket{1}+\ldots+\ket{2^n-1}$, in only $n$ steps, i.e., efficiently. Moreover, our ring cavity setup achieves this with a constant small number of optical devices.  
Transformation (\ref{inputtrafo}) thus yields the input state for the central transformation (\ref{eq:transformation}) of the Deutsch-Jozsa algorithm above. 
Instead of encoding the information about the function  $f$ separately, cp.\ transformation (\ref{eq:transformation}),  we combine it with the efficient preparation of $n$ qubits (\ref{inputtrafo}):
\begin{equation}
\ket{\psi_{\text{in}}} \rightarrow \sum_{x=0}^{2^n-1} (-1)^{f(x)}\ket{x}\,.
\label{trafo}
\end{equation}
A subsequent Fourier transform by means of a thin lens interferes the electric fields in the focal point (see \ref{app:fourier})  resulting in zero intensity for balanced functions and a finite intensity for unbalanced ones,  including constant functions. The intensity is a measure of how biased the function is, but this is not relevant for the Deutsch-Jozsa problem. 

It is  in principle possible to realise $(\ref{trafo})$ for any of the $2^N$ binary function $f$ with $N=2^n$ arguments by applying only phase shifts to beams propagating to the right, cp. Fig.\ \ref{fig:DiagEfPhase}. Alternatively,  $(\ref{trafo})$ can be imposed using phase shifts combined with controlled bit flips, cp. Fig.\ \ref{fig:DiagEfCnot}.  Normally, encoding the information about $f$ as in transformation (\ref{eq:transformation}) is outsourced to an oracle, i.e.\ a  black box. Here we take the standpoint that it must be possible to encode $f$ efficiently, otherwise the algorithm does not yield a result for a large number $N$.
To encode constant functions in our scheme is trivial. In case $f(x)=0$ for all arguments $x$, the transformation $(\ref{trafo})$ is identical to transformation (\ref{inputtrafo}). A global phase factor of $(-1)$ appears in case  $f(x)= 1$ for all $x$, but can be omitted because it is physically irrelevant.

On the other hand, to generate transformation $(\ref{trafo})$  for the $N \choose N/2$  balanced functions is in general non-trivial. However, our scheme allows to implement  balanced functions which correspond to product states efficiently by choosing  whether to apply to each newly generated qubit a relative phase shift:  
\begin{align}
&\sum_{x_1, \ldots, x_n= 0,1} (-1)^{f(x_1,\ldots, x_n)}\ket{x_1,\ldots, x_n}, \nonumber\\
&= \pm(\ket{0}\pm\ket{1}) (\ket{0}\pm\ket{1})\cdots  (\ket{0}\pm\ket{1}).
\label{trafo2}
\end{align}
This class of functions comprises $2^{n+1}$ elements corresponding to all possible combinations of relative phase shifts in $n$ qubits as well as a global phase shift. A small fraction of all balanced functions $f$ corresponding to entangled states  $\sum (-1)^{f(x_1,\ldots, x_n)}\ket{x_1,\ldots, x_n} $ can be programmed efficiently in our setup by the help of phase masks and controlled bit flips. Moreover, a relabeling scheme yields all balanced functions carrying out the algorithm without changing the physical realization. The relabeling procedure can depart from any state with equal distribution of $+1$ and $-1$ phase factors over the localised light fields and implements a desired balanced function by suitably changing the labels of the light fields.  As the renaming does not have any physical implication, the subsequent Fourier transform of the electric field yields zero intensity on the optic axis in the focal plain indicating a balanced function. Therefore, the renaming can be omitted, and any balanced function of $2^n$ arguments can be programmed efficiently by applying a single relative $\pi$ phase shift on the first of $n$ path qubits.

\section{Realisation Scheme}

\begin{figure}[htp]
\centering
\includegraphics[width=70mm]{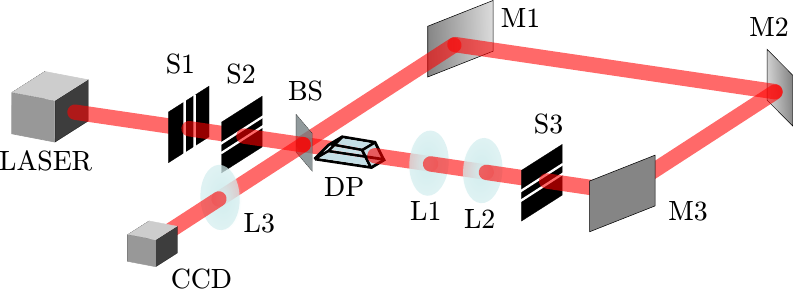}
\caption{Experimental implementation. S1--S3: slits; BS: beam splitter; DP: dove prism; L1 and L2 cylindrical lenses; L3 Fourier transforming lens; M1-M3: mirrors.}
\label{fig:Jozsa} 
\end{figure}

Our implementation scheme to generate the superposition  of $2^n$ states (\ref{inputtrafo}) efficiently, based on Dragoman's approach \cite{Dragoman2002}, is shown in Fig.\ \ref{fig:Jozsa}. A particular feature of our implementation is that it is based on classical light fields. One should note that the characteristic ingredients of quantum computation are present not only in quantum mechanics but also in classical optics. In particular, superposition, interference and a classical analog of entanglement \cite{Spreeuw1998, Luis2009, Simon2010, McLaren2015,Aiello2014} can be produced using coherent laser sources and linear optical elements. Classical optics has been employed to simulate quantum gates \cite{Dragoman2002}, to realise quantum walks \cite{Goyal2013b, Goyal2015} as well as to implement the Deutsch Algorithm \cite{PerezGarcia2015} and the Deutsch-Jozsa Algorithm \cite{Puentes2004} with classical states of light.

\begin{figure}[htp]
\centering
\includegraphics[width=60mm]{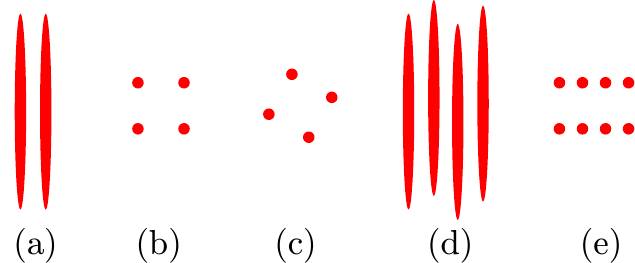}
\caption{Output after after: (a) S1, (b) S2, (c) GP, (d) L2, (e) S3.}
\label{fig:explain} 
\end{figure}

In the implementation, a qubit is created by a pair of slits that are illuminated by a pulsed laser source. These slits can be opened, closed or covered partly by a phase plate. For instance, to prepare a superposition of the form $(\ket{0}+\ket{1})/\sqrt{2}$ both slits are opened. The state $(\ket{0}-\ket{1})/\sqrt{2}$ can be prepared by covering one slit of a pair with a $\pi$ phase plate. In order to understand the principle of the implementation scheme, let us consider the first round trip inside the ring cavity. We assume that all  slits are open. Due to the combination of the slits S1 and S2, four distinct light dots  are created as indicated in Fig.\ \ref{fig:explain}-b. The transversal light field and thereby the four dots  are rotated by a Dove Prism (DP) as shown in Fig.\ \ref{fig:explain}-c. Then the light field is passed through a cylindrical lens system L1 and L2. The combination of these lenses produces four light stripes as depicted in Fig.\ \ref{fig:explain}-d. Finally, another pair of slits (S3) generates 8 light dots, as sketched in Fig.\ \ref{fig:explain}-e, by intersecting the four stripes. These steps can be repeated in order to create more dots. Note, that in each round trip the number of dots doubles. Furthermore, in order to minimise the overlap between dots, the distance between adjacent dots can be halved by changing the orientation of the DP. In addition, this procedure keeps the illuminated area of the light field on a transversal plane after each round trip constant  (see \ref{app:finite}).

The binary function $f$ is programmed in the presented setup as follows. For the constant functions $f(x)=0$  the setup shown in Fig.\ \ref{fig:Jozsa} is used without any additions. For the alternative constant function,  $f(x)= 1$ for all arguments $x$, the same setup can be used, since a global phase factor of $-1$ of the electric field, does not make any detectable difference.    For the balanced functions the setup is augmented by a  $\pi$-phase in front of a one slit of a pair of slits. By adding more phase plates in front of other slits or in particular round trips, it is possible to construct all balanced functions that belong to product states (\ref{trafo2}). Finally, the output of the cavity is Fourier transformed \cite{Goodman1996} by a lens (L3) and captured in a CCD camera. Light detected in the focal point behind the lens implies that the programmed function is constant, otherwise,  if no light is measured the function is balanced (see \ref{app:fourier}).

The present implementation scheme determines whether a programmed function is balanced or constant, by integrating the electric field, which carries the functional values as phase factors $\pm1$,  over a transversal plane. This integration is accomplished by a thin circular lens which superimposes the light from the focal plane in front of the lens in the focal point behind the lens. Hence the result of the integration stays the same, even if part of the light is superposed already during its passage through the ring cavity before the lens. Therefore, the scheme tolerates overlap between light spots when they are generated. As a consequence, it is not necessary to adjust the separation of dots in each round trip by dynamically changing the orientation of the DP. In addition, the number of roundtrips and thus the number of qubits generated is not restricted by the need to avoid overlap between dots. Although, above a certain degree of  overlap of the spots, the  qubits cannot any longer be measured  individually, which is not necessary  in this context, they can in any case be prepared with individual relative phase shifts.

We note that if the light spots cannot be resolved because of their overlap,  the concept of relabeling mentioned earlier becomes ambiguous. However, this is irrelevant since the relabeling can be omitted as discussed above.

\section{Conclusions}

The main problems of optical simulations of quantum computation are pointed out in Ref.\ \cite{Spreeuw2001}: the exponential  growth of the number of optical devices with the number of qubits, the exponential growth in space (cross section of the light field) and the decrease of optical power. The present implementation scheme solves the first problem  by the repeated use of the same optical elements in a ring cavity. The increase of the cross section of the light field, which results from doubling the number of light dots, can be constrained by either (i) adopting the orientation of the DP after each round trip (\ref{app:finite}) or by (ii) the use of a lens system \cite{Sudarshan1985} to compensate the increase in width of the pattern of dots without changing its height, the where the latter will compensate diffraction.  Finally, the optical power loss due to the filtering by the slits, might be addressed by using  a pair of cylindrical lenses which can reduce each light stripe to a pair of spots. Alternatively, since the scheme uses classical states of light,  losses can in principle be compensated by amplification inside the cavity.

In this work, we proposed a new way to implement the Deutsch-Jozsa Algorithm by using classical light. Our scheme makes use of linear optical elements inside a ring cavity to efficiently solve the task. We are able to create $2^n$ states in $n$ steps while encoding the functional values in the phase of the field. An experimental realisation of the scheme will give an indication of the extend of the domain of functions that can be tested. If the problem of losses can be solved satisfactorily, we expect to exceed the capacity of the Touring machines available at the time of publishing the present results. \vspace{10pt}

\section*{Acknowledgements}
We are grateful for discussions to R. Simon and K. Mpofu. This work is based on the research supported by the National Research Foundation Grant 93102. B.P. and R.I.H. acknowledge support from the Consejo Nacional de Ciencia y Tecnología (grant 158174).

\appendix

\section{Finite length of $2^n$ states}\label{app:finite}
By adapting the orientation of the Dove prism in the setup, the number of light spots can be doubled in such a way that the minimal distance is halved, starting with initial distance $d$ and leading to total length $L$ of 
\begin{equation}
L= 2^{n+1} d/2^{n} = 2 d
\end{equation}
 which is constant in all round trips $n=1,2\ldots$
The condition for the rotation of the square pattern of spots in Fig \ref{fig:explain} (c)  to  Fig \ref{fig:explain} (d) is given in terms of the rotation angle $\phi$ about an axis through the centre of the pattern:
\begin{equation}
\tan\phi= \frac{d}{d/2^n}=\frac{1}{2^n}\,.
\end{equation}
The dove prism must thus be oriented by the angle $\phi/2$ with respect to its longitudinal axis.

If we consider the finite size of the spots we have the following relation
\begin{equation}\label{eq:SizeDot}
n = 1 + \log_2{(d/\delta)},
\end{equation}
where $n$ is the iteration inside the cavity, $d$ is the initial separation between spots and $\delta$ is the size of the dots. 

\section{Fourier transform}\label{app:fourier}
The definition of the two dimensional Fourier transform of the complex function $g(x,y)$ reads
\begin{equation}
\mathcal{F}\{g(x,y)\} = \iint_{-\infty}^{\infty} g(x,y) \exp{[-i2\pi(ux+vy)]}\:\text{d}x\: \text{d}y,
\end{equation}
where $u$ and $v$ are generally referred as the frequencies in  Fourier space. Notice that if we evaluate at $u=v=0$, the Fourier transform is equivalent to 
\begin{equation}
\mathcal{F}\{g(x,y)\} = \iint_{-\infty}^{\infty} g(x,y) \:\text{d}x\: \text{d}y,
\end{equation}
which is the integral of the function $g(x,y)$ over the  $x-y$ plane. Furthermore, it is well known that a lens can efficiently perform a two dimensional Fourier transform \cite{Goodman1996}. This means that using a lens and measuring at the origin $(u=v=0)$ defined by the optical axis in the focal plane behind the lens, we obtain the integral of the field entering the lens from the focal plane in front of the lens.

\bibliographystyle{elsarticle-num} 
\biboptions{sort&compress}

\end{document}